\documentclass[10pt,journal,cspaper,compsoc]{IEEEtran}

\usepackage{url}
\usepackage{paralist}
\usepackage[usenames]{color}

\usepackage{Sweave}
\begin{document}
\Sconcordance{concordance:paper.tex:/home/azureuser/recomputation-ss-paper/SummerSchoolPaper/paper.Snw:%
1 9 1 1 0 86 1}

\title{Case Studies and Challenges in Reproducibility in the Computational Sciences}

\author{
Sylwester~Arabas,
Michael~R.~Bareford,
Lakshitha~R.~de~Silva,
Ian~P.~Gent,
Benjamin~M.~Gorman,
Masih~Hajiarabderkani,
Tristan~Henderson,
Luke~Hutton,
Alexander~Konovalov,
Lars~Kotthoff,
Ciaran~McCreesh,
Miguel~A.~Nacenta,
Ruma~R.~Paul,
Karen~E.~J.~Petrie,
Abdul~Razaq,
Dani\"el~Reijsbergen
and~Kenji~Takeda.%
\IEEEcompsocitemizethanks
{
\IEEEcompsocthanksitem S.~Arabas is with the University of Warsaw. M.~Bareford,
L.~de~Silva, I.~Gent, M.~Hajiarabderkani, T.~Henderson, L.~Hutton,
A.~Konovalov and M.~Nacenta are with the University of St Andrews.
B.~Gorman and K.~Petrie are with the University of Dundee. L.~Kotthoff is with Insight Centre for Data Analytics. C.~McCreesh is with the University of Glasgow. R.~Paul is with the Universit\'{e} catholique de Louvain. A.~Razaq is with Glasgow Caledonian University. D.~Reijsbergen is with the University of Edinburgh. K.~Takeda is with Microsoft Research.\protect\\
E-mail: emcsr2014@cs.st-andrews.ac.uk
}%
}

\IEEEcompsoctitleabstractindextext{
\begin{abstract}
This paper investigates the reproducibility of computational 
science research and identifies key challenges facing the community today. 
It is the result of the  First Summer School on Experimental Methodology 
in Computational Science Research.


First, we consider how to reproduce
experiments that involve human subjects, and in particular how to
deal with different ethics requirements at different institutions.
Second, we look at whether parallel and distributed computational
experiments are more or less reproducible than serial ones. Third,
we consider reproducible computational experiments from fields outside
computer science. Our final case study looks at whether
reproducibility for one researcher is the same as for another, by
having an author attempt to have others reproduce their own,
reproducible, paper.
This paper is \emph{open}, \emph{executable} and \emph{reproducible}: the whole process of writing this paper is
captured in the source control repository hosting both the source of the paper, supplementary codes and data;
we are providing setup for several experiments on which we were working;
finally, we try to 
describe what we have achieved during the week of the school in a way that others may reproduce (and 
hopefully improve) our experiments.
\end{abstract}
}

\maketitle
\IEEEpeerreviewmaketitle

\newcommand{\groupsubsec}[1]{\paragraph*{\textbf{#1}}}
\newcommand{\groupsubsubsec}[1]{\paragraph*{\textbf{#1}}}

\section{Introduction}
\label{s:intro}

In this paper we consider reproducibility in the computational sciences. We
interpret the computational sciences as including computer science, but also any
other science in which computational work plays an important role, such as
physics, biology or psychology. The central hypothesis of this paper is that
reproducibility is a cornerstone of the scientific method, and should therefore
be a cornerstone of the computational sciences. We ask to what extent this is true
and what challenges arise in reproducing work in the computational sciences.

To study these questions, we perform four case studies that consider different
issues related to reproducibility in the computational sciences.  Each case study
takes on a different aspect of reproducibility, identifies any problems
encountered, and discusses the points that they have raised.

An often-overlooked fact is that reproducibility in the computational sciences
is not only an issue of the actual computations, but also of preliminaries and
other setup that is not related to computation at all. The first case study asks
how reproducibility is affected by differing ethics requirements at different
universities. Ethics approval has to be obtained for any experiments involving
human subjects, such as the ones that are common in human-computer interaction
research, but different institutions put emphasis on different aspects. How
reproducible is research that is ``conditioned'' on one particular ethics form?

The second case study considers whether parallel and distributed computational
experiments are more or less reproducible than serial ones. Computational
experiments are at the core of much research, such as in artificial
intelligence, where the motivation for this case study lies. In recent years,
multi-core and multi-processor machines have become increasingly prevalent. To
make use of this increased processing power, experiments need to utilise
multiple resources at once. But is this inherently detrimental to
reproducibility?

While computer science may account for a significant fraction of computational
experiments, they are important in almost every scientific discipline. In environmental science, 
medicine, physics and chemistry for example, simulations of physical processes help scientists gain
insights. Our third case study is motivated by this fact and asks how
reproducible computational scientific experiments are from disciplines other than
computer science. We consider experiments from several research areas. Given
that researchers in other areas use computation only as a tool and may not be
as aware of issues related to reproducibility as computer scientists, does
reproducibility suffer?

Our final case study considers whether reproducible for one person is the same
as reproducible for another. It asks how reproducible is the data analysis of a
published experiment in human-computer interaction where the author made
significant efforts to make it reproducible. While it is reasonable to assume
that the author of an experiment can conduct it in a way that enables her or him
to do it again, having another person do it is an entirely different matter.
There may be implicit assumptions that are not specified, background knowledge
assumed, or environmental aspects unconsidered. Can a carefully prepared
experiment be reproduced by someone with no specific background in the area?

Most of the work for this paper was performed during the Summer School on
Experimental Methodology in Computational Science Research.\footnote{St
Andrews, Scotland, August 4-8, 2014,
\url{http://blogs.cs.st-andrews.ac.uk/emcsr2014/}}
Indeed, a highly
provisional first draft of this paper was completed by the end of the
school~\cite{emcsr_arxiv_draft}, with the ensuing weeks used in completing work and writing.  
The first three case studies
were selected by discussing the interests of participants and forming groups
and research questions around these. Therefore, these case studies were
performed by subsets of the authors, and are
presented below in Sections~\ref{s:group1}~--~\ref{s:group3}. The final case
study was led by a lecturer of the summer school, was performed by all
participants, and is presented in Section~\ref{s:group4}.

The case studies we consider are not exhaustive and no single one can give
complete answers. Instead, they shine spotlights on specific, important areas
related to the issue of reproducibility in the computational sciences. Furthermore,
each raises interesting questions for future consideration by researchers
interested in the reproducibility of computational experiments, be it trying to
reproduce someone else's experiments, or making their own experiments
reproducible.

Additionally, we consider the meta-level problem of how to make this very paper
reproducible. We have striven to make this paper open, reproducible, and
executable.
The entire edit history, including paper and many aspects of the
case studies, is available openly on
GitHub
and was indeed open from the start
of writing the paper~\cite{summerschoolpaper}.
The paper has executable aspects through the integration of R
and \LaTeX\ via the R package Sweave~\cite{lmucs-papers:Leisch:2002}, so that as the underlying
data change, new tables and figures can be regenerated automatically. Finally,
we have endeavoured to make it reproducible through measures such as providing
virtual machines (VMs) for aspects of our work, not least a VM in which
the paper itself can be rebuilt, and we welcome readers to attempt to
reproduce it - either locally or using cloud computing.

\section{The state of the art}
\label{s:recomputation}

Reproducibility, replicability and the like have been discussed for
many years~\cite{lubin:replicability} and are acknowledged as a
fundamental part of the scientific method. More recently, interest has
arisen in reproducibility specifically in the computational sciences, with
some considering the nature of the computational sciences to make the
issues distinct from other sciences~\cite{donoho:reproducible}.
Despite its importance, various studies have shown that many, if not
most, research papers are not
reproducible~\cite{bonnet:repeatability,hornbaek:replications,ioannidis:repeatability}.
This has led to calls for independent boards to replicate and certify
research experiments~\cite{baker:verify}. At the same time, efforts
are ongoing to improve methodologies into determining whether an
experiment has been reproduced accurately~\cite{johnson:evidence}.

Terminology is important, but unfortunately many terms are used
interchangeably in this area.  Peng~\cite{peng:reproducible} describes
a spectrum of reproducibility from a paper that is not reproducible,
to one that allows ``full replication''. But
Drummond~\cite{drummond:replicability} distinguishes between
replicability and reproducibility, by saying that the former is the
exact repeating of an experiment as presented, whereas reproducibility
is broader and allows one to build on an experiment and further
science. Gent introduces another term, \emph{``recomputation''}, to describe
the replication of computational experiments (The Recomputation
Manifesto~\cite{gent:recomputation}).
In this paper we consider challenges and ways in which we can enable
the widespread recomputation of experiments as described in the
Recomputation Manifesto.

There are many aspects to recomputation, all of which are being actively
studied. Stodden and Miguez propose a set of best practices for reproducible and
extensible research, including licensing and sharing of data, workflow tracking,
making code and method available, and citing data and
software~\cite{stodden:practices}. Companies such as
FigShare~\cite{figshare:citable} and GitHub~\cite{github:citable} are helping by
making it easier for researchers to share code and data and also cite them
through the use of DOIs (Digital Object Identifiers). Several initiatives have
arisen to make researchers more aware of tools, and to further the development
of tools, to make their research more reproducible, e.g., Mozilla Science
Labs~\cite{mozillasciencelabs},
Open Science Framework~\cite{openscienceframework},
RunMyCode~\cite{runmycode}
and Software Carpentry~\cite{softwarecarpentry}.
Davison describes how
to make it easier to capture workflow and experimental context, including using
a ``consistent, repeatable computing environment'' (the recomputation.org
project and others~\cite{howe:reproducible} aim to make this particular aspect
easier through the use of VMs), version control and clearly
separated experimental parameters~\cite{davison:reproducibility}. Mesirov
describes one particular workflow to make it easy to track and package genomic
data~\cite{mesirov:accessible}. Our paper attempts to follow these best
practices, uses version control and is available to recompute on a VM either 
locally or in the public cloud.

\section{Obstacles to Reproducibility in the Computational Sciences}
\label{s:obstacles}

There are various obstacles to reproducibility, and many of these were
  discussed by speakers and participants during our summer school.
The obstacles and challenges are related to different stages of scientific 
  endeavour and different aspects of reproducibility.
Amongst the obstacles and challenges that we identified are:

{\bf Incentives} -- even though, as outlined in Section~\ref{s:recomputation}, 
  the issue of reproducibility is
  currently actively discussed by the scientific community, 
  and funding bodies do campaign for
  openness in the computational sciences (e.g.,~\cite{NSF}), 
  the prevalent impression from the discussions at 
  the summer school is that there is yet a lot to be done to increase 
  awareness of how one can benefit from reproducibility.
Mechanisms that reward those who deliver reproducible results are also
  still to become widespread.
It was reiterated during the school that the self-benefits 
  of reproducibility, such as increased productivity, are likely 
  a good enough reason to invest time on assuring it.
But both the benefits to the original researchers and the benefits to the community
  are latent and perceptible only in the longer term.
Thus investing the time to offer reproducibility is difficult without
  approbation and endorsement from colleagues, collaborators, supervisors,
  reviewers and executives.
  
{\bf Prerequisites} -- offering reproducibility in computational
  science is linked with giving access to data and code, both of which may
  be subject to legal limitations due to intellectual property and
  data protection issues. Indeed we encountered some of these issues
  in one of our case studies (Section~\ref{s:group3}).
Sharing data may additionally be bound to ethical responsibility, for instance
  if the data contain personal records or other sensitive information. 
Even non-sensitive non-commercial data may not be readily redistributed
  without assuring proper ethical consent from the data originators.

{\bf Practicalities} -- there is definitely no consensus as to the
  best technique for facilitating reproducibility of computational
  experiments. 
The reasons are at least threefold. 
First, the wealth of nomenclature (replicate, recompute, reproduce, 
  rerun, repeat, reuse) indicates that the aims and the needs of researchers
  differ significantly.
Second, technologies advance and continuously offer new methods that can
  aid researchers in offering reproducible results.
Third, there are numerous tradeoffs to be made: 
  \begin{inparaenum}[(i)]
  \item embrace cutting edge or widespread technology; 
  \item offer standalone or integrable solutions; 
  \item prioritise independence or ease of reproduction; 
  \item disseminate reduced comprehensible datasets or avoid excluding any data;
  \item put efforts on automating experiments with already publishable results
  or move on to new experiments.
  \end{inparaenum}
    
{\bf Dissemination} -- even if all of the above obstacles are overcome, 
  distribution and long-term persistence of reproducible experiments
  pose challenges.
First, finding a venue for long-term preservation of large datasets or even 
  moderately-sized VM-based packages may be an issue, especially
if one aims at blending software and data
  dissemination with traditional academic publishing.
Second, given that one of the key aims of reproducibility in science
  is to offer independent verification, the amount of knowledge 
  needed to rerun an experiment has to be reasonably adapted to the
  target audience.
Interdisciplinarity is inherent in the idea of the computational
sciences, which
  merge computational techniques with diverse domains of science.
Consequently, one has to take into account that the users who will attempt to
  reproduce the results of an experiment may be from a different discipline.
Last but not least, dissemination of data, and especially of executable
  programs is subject to security issues associated with both the security
  of computer systems and the safety of the personal data of the researchers who
  are preparing the experiment and reproducing the results.

\section{Case study \#1: Ethical requirements for recomputation}
\label{s:group1}

Human subjects research involves the collection of data through
interaction with individuals, or through collection of personally
identifiable information. Such research poses specific barriers to
reproducibility. In fields such as human-computer interaction (HCI),
there are few recomputations of previous work, attributable to a
culture that does not reward reproducibility, difficulty in
replicating interaction techniques when materials are not shared, and
an emphasis on formative work which proposes new techniques over
summative work~\cite{hornbaek:replications}.

In the first of our case studies, we look at one specific challenge to reproducibility in HCI
research: capturing and disseminating the ethical requirements of an
experiment, such that others may better recompute the procedures of a
study. What we wish to make possible is the following scenario: 

\renewcommand{\labelenumi}{Step \arabic{enumi}:}
\begin{enumerate}
\item Alice undertakes an experiment where she collects human data. 
\item Alice then analyses those data and publishes an academic paper.
\item To comply with recomputation standards Alice then creates a VM 
which contains the data she collected and scripts to create the statistics
used in the paper. She includes an ethics specification which enumerates
key methodological details and ethical considerations,
and places the VM on recomputation.org. 
\item Bob reads Alice's research paper and is interested in comparing her work to his own.
\item Bob goes to recomputation.org and downloads the VM.
He recomputes all of Alice's experiments in order to verify the analysis in her paper.
\item Bob compares the data Alice has provided with his own and publishes his own paper.
From Alice's ethics specification, he can directly compare ethical considerations to account
for any methodological differences.
\item To comply with recomputation standards Bob then creates a VM which contains 
the data he collected and Alice's data and scripts to create the statistics used in the paper. He includes his own ethics specification which can be directly compared to Alice's for
the benefit of any further recomputations, 
then places this VM on recomputation.org.
\end{enumerate}
\renewcommand{\labelenumi}{\arabic{enumi}}

We evaluate the ethics requirements procedures of ten universities to
determine a minimum specification for reporting ethical considerations.

\groupsubsec{4.1 Background}
Human subjects research often involves the collection
of sensitive identifiable data about participants. To ensure participants are
not placed at undue risk by the conduct of an experiment, an increasingly
rigorous process of oversight of the ethical conduct of research institutions
has emerged in recent decades in many countries, particularly the US, where
institutional review boards (IRB) have been charged with reviewing all
clinical and human subjects research in line with federal regulations, with
significant penalties for institutions if ethics violations occur. Each
institution, however, has freedom to implement IRB processes as they see fit
so long as such regulations are upheld. This leads to great inconsistencies
between the expectations of institutions, and the processes researchers must
engage with in order for research protocols to be approved.

Internationally, the situation is even more variable. In the UK, research
councils mandate that ethics be considered in order to receive funding, but the
conduct of individual ethics committees is not regulated. Some countries may
not impose any requirements at all upon institutions.

Such variety in the conduct of ethical approval between institutions a
represents a significant barrier to reproducibility in HCI research. If a
researcher wishes to recompute a HCI experiment which uses human subjects
data, they will usually need to seek ethical approval from their own
institution. In our recent work, a survey of 505 papers using online social network
(OSN) data found that only 2\% of papers disclose any of the ethical
considerations of their work~\cite{hutton:reproducibility}.

As researchers do not routinely disclose the protocol that received
IRB approval, attempted recomputations may miss crucial details
necessary to conduct the experiment. With IRBs and ethics boards
operating largely independently with little policy coordination, there
is no standardisation of ethics procedures. A study in one institution
might be difficult to replicate elsewhere, if the IRB at the latter
could not interpret the original ethics application.

Yet, while reproducibility has only recently been considered an
important ambition for HCI researchers, with nascent efforts including
RepliCHI, which has operated as a CHI workshop since
2013~\cite{wilson:2013}, the wider community has not considered these
ethical challenges in detail.

We assess the state of the art in ethics procedures to determine what
commonalities exist between institutional requirements. From this, we aim to
derive a minimum ``ethical specification'', encoding fundamental
methodological details to help researchers recompute procedures and ethical
details, and to make it easier to replicate applications to other IRBs, with
the ambition of such specifications being routinely attached to HCI
experiments.

\groupsubsec{4.2 Ethical Requirements across Institutions}

To understand the state of ethics procedures between institutions, we
collected ethics applications forms for ten universities located in the UK,
EU, USA, and Asia. A range of locations were chosen and a mixture of both
large research intensive and smaller institutions to capture a range of
cultural and regulatory expectations, which we expect will manifest in
different procedures. All forms collected were from publicly
accessible sources, except for one supplied by the authors. This is in itself
a significant barrier to reproducibility. Without making procedures publicly
available, there can be no external scrutiny about an
institution's procedures, which makes it more difficult to derive standards.

For each form, two researchers independently identified unique fields, accounting for
differences in wording between forms so long as each attribute asked for the
same atomic information. Where one form requests expanded information
pertaining to a previously identified attribute, this was considered a
\emph{sub-attribute}. After independently coding the forms, the two researchers
discussed any discrepancies to arrive at a set of 145 unique attributes,
encompassing generic details, such as contact details of co-investigators,
methodological details, and institution-specific requirements, often for
insurance and liability purposes. Of these fields, only two were common to all
ten ethics forms --- the name of the principal investigator, and whether
informed consent was sought. This intersection was significantly smaller than
anticipated, and clearly does not constitute a useful minimum ethical
specification. It does however reveal two interesting properties. It confirms
our intuition that ethical procedures vary greatly between institutions, while
also identifying perhaps the single most important objective of the ethics
process: to ensure participants have given informed consent to participate in
an experiment.


\begin{figure}

\begin{minipage}{\linewidth}
\begin{center}
\includegraphics{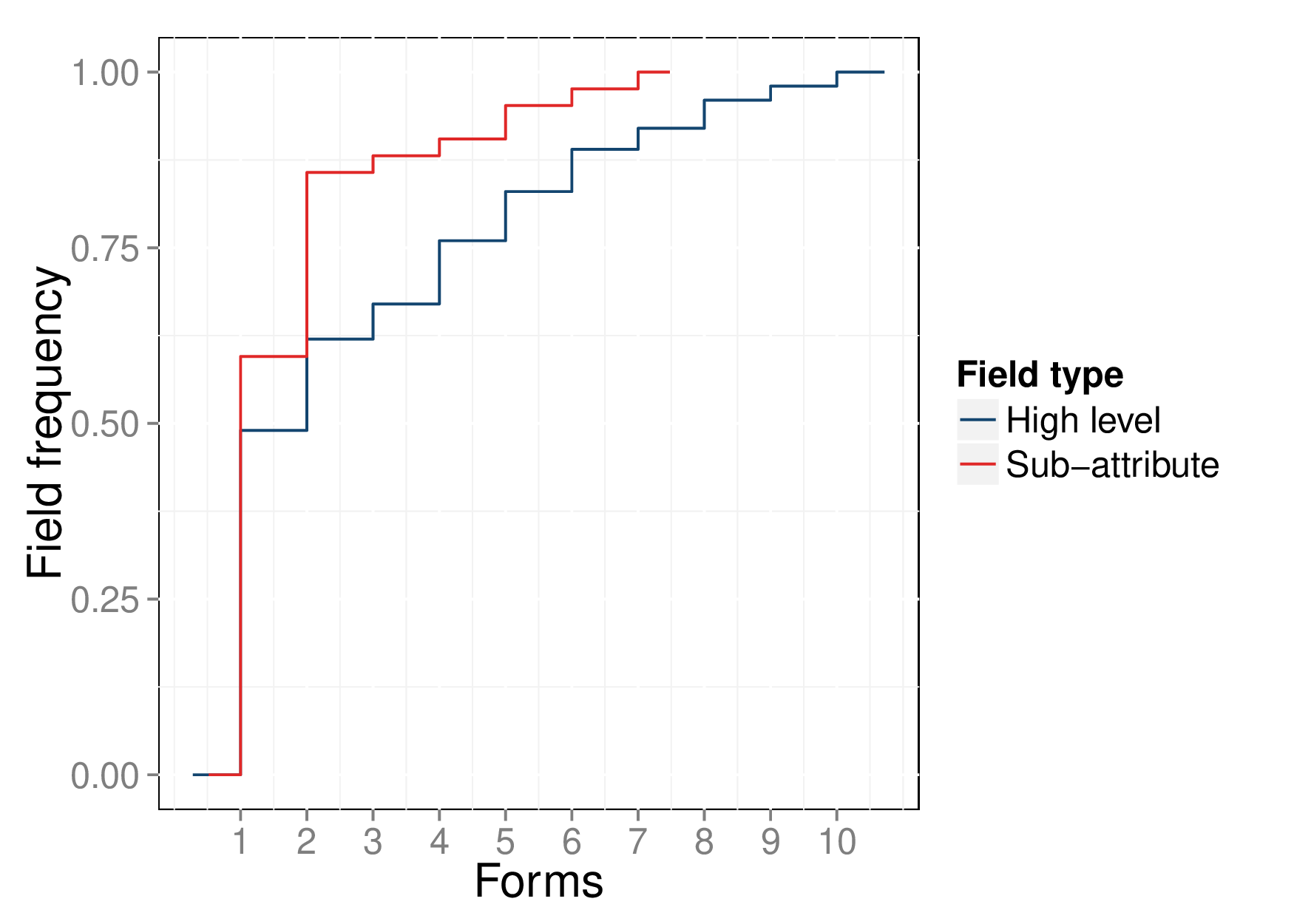}
\end{center}
\caption{\label{p:ethicsFreq}CDF showing the distribution of high-level and sub-attributes across the ethics forms examined. Half of high-level attributes only occur in one form, while no sub-attributes appear in more than seven forms.}

\end{minipage}
\end{figure}

Figure~\ref{p:ethicsFreq} shows the distribution of attributes across the ten
forms we examined for both high-level and sub-attributes. As shown, half of
high-level attributes only appear in one form, while 60\% of sub-attributes
are unique to one form. For example, while 60\% of forms ask whether
participants receive financial compensation (high level), only one asks
whether co-investigators are compensated (sub-attribute). We are most
interested in the unique high-level attributes that emerge, as it is important
to discern between questions which capture institution-specific requirements,
or may constitute important issues which other IRBs ought to consider. We find
instances of both in our results. For example, while UCL are the only
institution to ask whether their own students are participants in the research
(we assume for liability reasons), surprisingly they are the only university
to ask outright whether health and safety precautions have been considered.
Interestingly, only Aga Khan University in Pakistan asks whether the study is
a replication of a previous experiment.

\groupsubsec{4.3 Proposed ethics specification}
Given the small intersection of attributes in our study, we isolate the 20 
most common high-level attributes -- those which occurred in six or more of
the ten forms we examined. We combine any semantically similar fields to
produce the following set of 15 attributes, presented in descending order of
frequency.

\begin{itemize}
	\item Was consent sought?
	\item Was deception involved?
	\item Project title
	\item Study duration
	\item Are there risks to participants?
	\item Justify use of vulnerable participants
	\item PI contact details
	\item Funding body information
	\item Is likely to induce participant stress?
	\item Summarise research proposal/experimental methods
	\item Are supplementary documents attached? (consent forms, briefing info etc.)
	\item Are participants financially compensated?
	\item Is study clinical?
	\item Supervisor name
	\item Describe ethical issues 
\end{itemize}

This set of attributes covers a range of fundamental methodological details,
many of which can be encoded in a consistent fashion and attached as metadata
to support replications. In future work, we aim to demonstrate whether this set
of attributes is sufficient to capture key methodological details, which we
cannot assert from this strictly frequency-based exercise.

\groupsubsubsec{4.3.1 Limitations}
This analysis is not intended as a rigorous survey of ethics procedures
internationally. The selection of institutions is inherently biased, as we are
only able to extract forms which are publicly accessible, a barrier to
reproducibility, and we have a particular emphasis on the UK in this study,
with six institutions represented. The intent of this exercise is not to make
statistical inferences about the state of the art in ethics review, but rather
to motivate the minimum set of attributes we recommend researchers disclose
when sharing their experimental methodologies. We also wish to raise awareness
of the importance of ethical procedures when considering reproducibility of
research.

\section{Case study \#2: Recomputing parallel and distributed experiments}

The Recomputation.org project has shown that recomputation via a 
VM can be successful when using sequential code on a single
machine.  Our second case study examined 
whether this approach could be extended to cover multi-core
parallel or distributed systems. For parallelism research, the end goal is
typically performance: can using more cores make something run faster, or allow
a larger dataset to be processed in the available time? It is not clear whether
a virtualised environment would affect the quality of the results for these
kinds of experiments. For distributed systems, a single VM would
obviously not suffice, but could multiple VMs be used?  To address
these questions, we looked at a multi-core parallel experiment, and a
peer-to-peer system.

\groupsubsec{5.1 Experiment 1: A Parallel System}

We looked at an existing implementation~\cite{parasols}
of a parallel branch-and-bound
algorithm for the maximum clique problem~\cite{ciaran}: given a graph, a clique
is a subset of pairwise-adjacent vertices, and the maximum clique problem
(which is NP-hard) is to find the largest such subset. On physical hardware, for
any given graph instance, if we run the program multiple times we get very
similar runtimes.  The aim of this experiment is to see whether runtimes are
similarly consistent when run on a VM on a public cloud computing infrastructure: the original
work relies upon performance measurements to explain the behaviour of different
parallelism mechanisms. We did not attempt to reproduce the entire paper;
instead, we devised a new experiment to establish whether meaningful,
consistent parallel performance measurements of the kind required by the
original work could be obtained using the Microsoft Azure cloud~\cite{azure}.

The source code of the implementation was available on GitHub (and was written
by one of the authors). No changes were required, and all of the dependencies
were available pre-packaged on a standard Ubuntu installation.

For our analysis, we selected $10$ different ``medium sized'' problem instances
(i.e., graphs) from the second DIMACS implementation
challenge~\cite{dimacs}, so that our
runtimes would be long enough to be noise-free but short enough to be
repeatable. This is a publicly available dataset, in an easy to parse format,
which is widely used for testing maximum clique algorithms, and there are no
barriers to obtaining or using it.

For each selected instance, we ran our implementation of the algorithm $50$
times on physical hardware (using $4$ cores of a machine with dual Intel Xeon
E5-2640 v2 processors), and then $50$ times on a VM with $4$ cores on Microsoft
Azure. Our measurements include only computation times and ignore the time
taken to read in the problem instances from a file. We compare the coefficient
of variation of the runtimes
on real and virtual hardware for each problem instance.  We present the results
in Table~\ref{run_time}. Each row of the table represents a problem instance.

\begin{table}[ht]
\centering
\begin{tabular}{lrr}
  \hline
 & real & vm \\ 
  \hline
brock400\_1 & 0.005 & 0.003 \\ 
  brock400\_2 & 0.007 & 0.002 \\ 
  brock400\_3 & 0.008 & 0.001 \\ 
  brock400\_4 & 0.010 & 0.003 \\ 
  MANN\_a45 & 0.005 & 0.002 \\ 
  p\_hat500.3 & 0.005 & 0.002 \\ 
  DSJC1000\_5 & 0.004 & 0.002 \\ 
  p\_hat1000.2 & 0.005 & 0.003 \\ 
  sanr400\_0.7 & 0.005 & 0.003 \\ 
  p\_hat700.3 & 0.004 & 0.002 \\ 
   \hline
\end{tabular}
\caption{Coefficient of variation of runtimes for different problem instances, when run on real or virtual hardware. In every case, the value is very low, implies virtualised hardware is as reliable as real hardware.} 
\label{run_time}
\end{table}
In both physical and virtual hardware the coefficient of variation is very small
(at or below $0.01$ in every case). We did not encounter any abnormalities when
running on a VM; this is contrary to the experiences of
Kotthoff~\cite{kotthoff}, who did not always see reliable \emph{sequential}
runtimes on
virtualised hardware. In other words, using virtualised hardware for
reproducible parallelism experiments is not necessarily infeasible.  However,
we were limited to 4 cores, rather than the 64 cores used in the original work.
We are also unsure whether parallel acceleration hardware such as GPUs or the
Intel Xeon Phi could be used in a virtualised environment, and are doubtful
that experiments involving this kind of equipment will be recomputable on later
hardware.

The VM image used for the experiments is available via
VMDepot.\footnote{\url{http://vmdepot.msopentech.com/Vhd/Show?vhdId=44545}}

\groupsubsec{5.2 Experiment 2: A Peer-to-Peer System}
Experiments that are performed across multiple machines are challenging
to reproduce. This is due to the cost of needed resources, and the complexities
involved in configuring the machines, the network, and the relationship between the machines.
Cloud services offer increasingly affordable computation. Such services use
virtualisation to decrease the cost of system (re)configuration.

In order to study the challenges of making distributed experiments reproducible
using cloud services, we deployed a Chord~\cite{chord} Distributed Hash Table
(DHT) of $10$ nodes on $10$ dedicated machines, and tried to reproduce the same
experiment on Microsoft Azure.

Reproducing the same experiment on Microsoft Azure proved to be time consuming 
and we could not reproduce the experiment during the Summer School. We were 
unable to automate a required network-specific port allocation task used in 
this experiment due to lack of time. It highlights the potential complexity of cloud 
computing APIs when being applied in research experiments, as opposed to 
more common deployment scenarios that have more examples and documentation.

It is possible to run experiments across multiple machines reproducible by writing
a vendor-specific script that starts and configures any needed VMs before running
the experiment. This approach raises a number of issues:
\begin{inparaenum}[(i)]
\item it relies on external
services in order to run the experiment,
\item it is time-consuming to produce such an
script, and
\item the script cannot be re-used on other cloud services.
\end{inparaenum}

The use of vendor-agnostic cloud frameworks such as Docker and LibCloud, may 
provide additional flexibility when deploying to different cloud infrastructures. Further 
studies are needed to identify best practices to make distributed experiments reproducible 
using VMs. 

\section{Case study \#3: Recomputing non-CS experiments}
\label{s:group3}

In this case study we focus on the issues relevant to research reproducibility that are potentially 
  unique to non-CS computational research.
The discussion is based on experience gained while packaging three computational experiments 
  into recomputable VMs.
The papers on which the discussion is based 
  deal with urban planning~\cite{danielpaper}, solar
  physics~\cite{bareford2010nanoflare}, and atmospheric
  physics~\cite{arabas2013libcloud},
  although the discussion is likely relevant to other non-CS domains. One of the
  authors of each paper was involved in this case study.

\groupsubsec{6.1 Methodology}

Our aim was to offer potential readers (or reviewers) of the papers in question, 
  the opportunity to reproduce the figures presented in the papers, 
  to inspect the code, and to possibly test behaviour of the programs with other parameters.
To offer it all within a ready-to-use environment, we have used the Vagrant
tool\footnote{\url{https://www.vagrantup.com/}}
to
  construct a single VM with all needed software and its dependencies installed.
The VM was based on the Debian Sid GNU/Linux distribution.

\groupsubsec{6.2 Results}
\groupsubsubsec{6.2.1 Experiment 1: Urban Planning}

For the first paper, we tried to recompute parts of Table~6 of
\cite{danielpaper}, in which the punctuality of a bus service in Edinburgh is
evaluated using statistical methods. Table~6 contains confidence intervals that
are constructed using a piece of software written in Java. It is possible to
compile the source code after installing a Java Software Development Kit on the
VM. The software uses the SSJ library for Stochastic
Simulation~\cite{ssj},
which had to be downloaded from the Internet. After that, we were able to run
the code.

We encountered legal obstacles in two areas.
\begin{itemize} 
\item Code-related: the software was developed as part of the EU project
    QUANTICOL~\cite{quanticol}, which is funded by the European Commission as part of its 7th
Framework programme. In the General Conditions part of the project's grant
agreement, it is stated that the IP rights are awarded to the beneficiary, which
in this case is the University of Edinburgh. In turn, the University of
Edinburgh has issued a position statement on intellectual property
rights~\cite{edinburgh:ipr}
in which it stated that it ``\emph{is the policy of the University of Edinburgh to develop University research capabilities and to assess, develop and promote the transfer of Edinburgh's technology and ideas for society's use and benefit.}'' Still, publication of the code would need to be checked with a supervisor.
\item Data-related: the code uses a dataset based on bus location measurements provided to us by Lothian Buses. We would need their permission to put this dataset in the public domain.
\end{itemize}
The use of the SSJ package is not an obstacle because it is released under the GPL licence from GNU.

The resulting VM contains the Java source files, allowing
researchers with knowledge of Java to analyse the correctness of the programme.
However, the code is not documented and may be hard to read. Specifically,
parameters are hard-coded and no interpretation is given of the produced
numbers and \LaTeX\ code.

The output of the Java code is displayed in Table~\ref{tab: z bootstrapped cis}.

\newcommand{\vtwo}[2]{$\left[\begin{array}{l} #1, \\ #2 \end{array}\right]$}
\newcommand{\exr}[1]{\cdot$10$^{#1}}
\newcommand{\hpi}[1]{\hat{\pi}_z(#1)}
\newcommand{\hpic}[1]{\multicolumn{1}{c}{$\hpi{#1}$}}
\newcommand{\hpicl}[1]{\multicolumn{1}{|c}{$\hpi{#1}$}}
\newcommand{\hpicr}[1]{\multicolumn{1}{c|}{$\hpi{#1}$}}

\def\arraystretch{1.2}
\setlength{\tabcolsep}{1pt}
\begin{table}[htbp]
\centering
\begin{tabular}{| c c c c c  |} \hline 
 \hpicl{5} & \hpic{6} & \hpic{7} & \hpic{8} & \hpicr{9} \\
 \vtwo{0}{0}&\vtwo{2.875\exr{-4}}{4.253\exr{-4}}&\vtwo{0.434}{0.485}&\vtwo{0.513}{0.566}&\vtwo{0.0}{0.0}\\[10pt]
\hline
\end{tabular}
\caption{A reproduction of the first row of Table 6 of~\cite{danielpaper}, which contains confidence intervals for estimates of $\pi_z(k)$. For each $k$, $\pi_z(k)$ denotes the probability that, upon arrival to the bus stop near Edinburgh airport, $k$ arrivals of buses of Route~100 are observed in the next hour.}
\label{tab: z bootstrapped cis}
\end{table}
\def\arraystretch{1.5}

\groupsubsubsec{6.2.2 Experiment 2: Solar Physics}


The work discussed in~\cite{bareford2010nanoflare} concerns the energy released
from an idealised cylindrical magnetic field when it becomes unstable and
subsequently relaxes to a simpler state. Initially, a field (or loop) starts in
a stable configuration and is then taken on a random walk through a known two
dimensional region of stability, see \mbox{Figure 3} of the aforementioned
publication. When the field crosses the boundary of this region (i.e.\ the threshold for instability), an energy release is determined and the field is moved to simpler stable configuration. 

When this process is repeated many times, the results can be expressed as energy distributions, see Figure~14 of~\cite{bareford2010nanoflare}, which shows the energy releases for $10^5$ relaxations involving a variety of loop lifetimes. We decided to focus on reproducing only this figure. The individual plots were taken from the results produced by a C++ code called Taylor Relaxation of Loop Ensembles, or TRoLE for short, that automates the process described above.

The TRoLE code was written when the first author of~\cite{bareford2010nanoflare}
was a postgraduate at the University of Manchester. It was necessary therefore
to check the IP policy for this
institution~\cite{manchester:ipr}.
Section~3.2 of this policy states that the ``\emph{...ownership of IP created by
a Student, who is not an employee of the University, is with the Student.}'',
which we took to mean that the first author was free to place the code in a
publicly available
repository~\cite{trole}.

The next issue concerned a technical matter: the code uses
a proprietary numerical algorithms library
(NAG)~\cite{nag}, the 
licence for which had expired. Fortunately, there is an open source alternative,
the GNU Scientific Library
(GSL)~\cite{gsl}. The TRoLE code was updated such that all NAG calls were replaced with the GSL alternative. 

%

\begin{figure}
\begin{minipage}{\linewidth} 
  \center
  \includegraphics{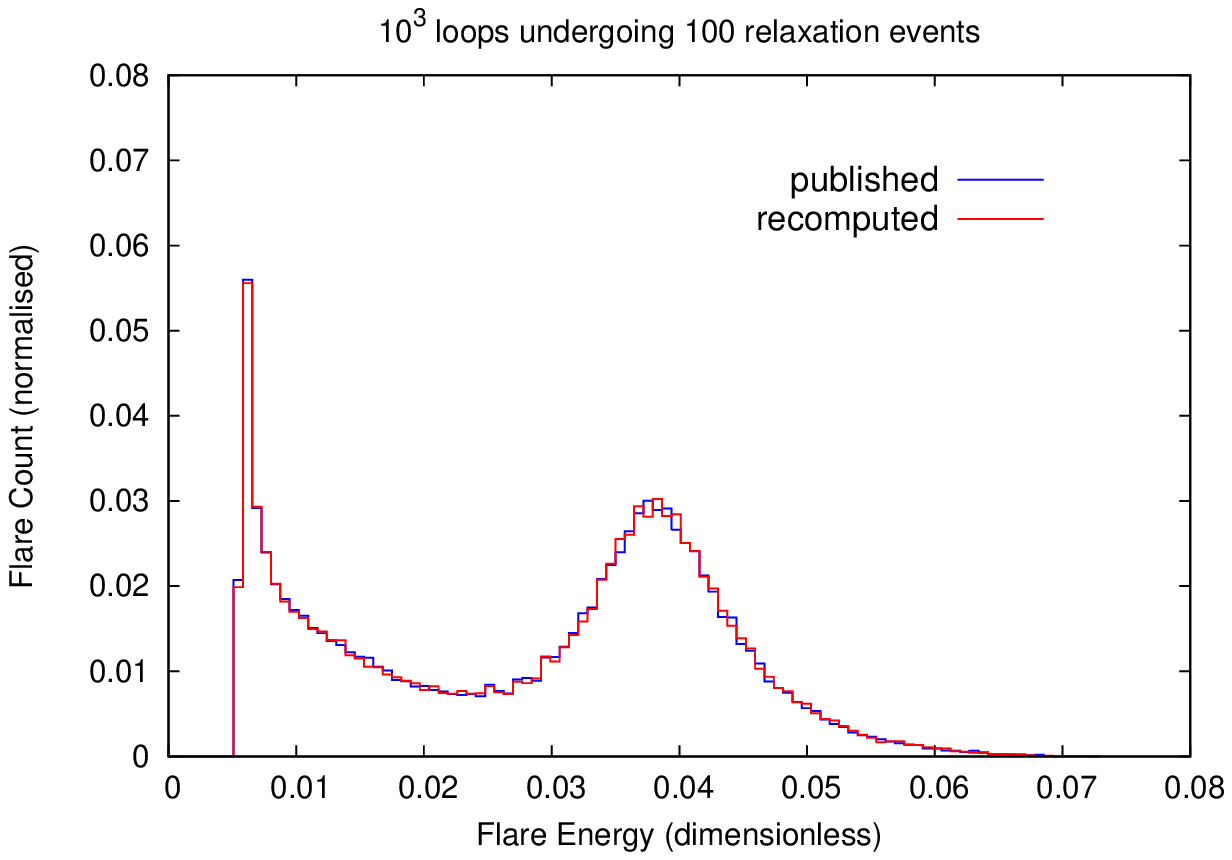}\\
  \vspace{5pt}
  \includegraphics[scale=0.001]{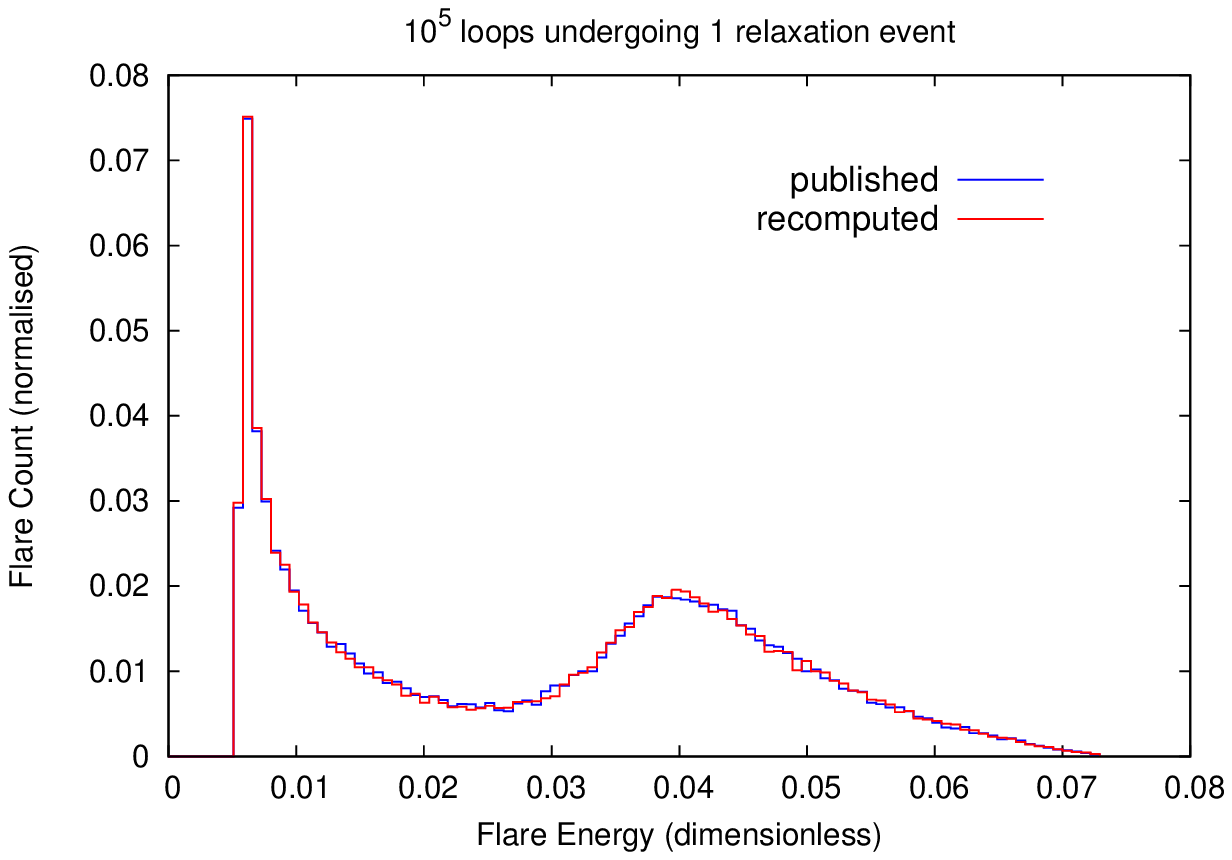}
  \caption{\small{Flare energy distributions over 10$^5$ relaxation events for two different loop lifetimes, 100 (top) and 1 (bottom) relaxation event(s). The red plots are the recomputed results and the blue are the results from the original paper~\cite{bareford2010nanoflare}.}}
  \label{fig_recomp_grp3_exp2}
\end{minipage} 
\end{figure}

Figure \ref{fig_recomp_grp3_exp2} shows two of the recomputed plots, the published energy distributions 
 ~\cite{bareford2010nanoflare} are given in blue.
The differences between the two sets of results are negligible and entirely consistent with the stochastic method used to generate the energy distributions.

\groupsubsubsec{6.2.3 Experiment 3: Atmospheric Physics}

For the third paper, we prepared pre-built software for reproducing 
  simulations of idealised atmospheric clouds presented in~\cite{arabas2013libcloud}.
The paper covers description of an open-source C++ library named libcloudph++. 
The library is intended for representing microphysics of clouds and precipitation
  in numerical models -- coupling libcloudph++ with a fluid-dynamics solver 
  of an atmospheric flow allows one to study the formation of clouds and rain.

While the software has already been released under the open-source GPL license
  and has been accessible through a GitHub repository, there is still
  a potential legal issue.
The library features the implementation of an algorithm inspired by a technique 
  described in the scientific literature but also covered by several patents,
  including a European one~\cite{shima2007simulation}. This prevents us from using the
  existing implementation.

Furthermore, the crux of the implementation is its support for execution
  on both CPUs and GPUs, the latter being optional but offering 
  significantly faster executions times.
However, the implementation uses the non-free CUDA standard what limits its potential
  users to owners of hardware of particular vendor and related proprietary software.

The only challenge in setting up the VM was in providing all needed
  dependencies, namely recent versions of a C++ compiler, and the CMake, Boost and Thrust 
  packages.

\groupsubsec{6.4 Discussion}

We encountered legal obstacles with all papers.
All of the software had been developed at universities, which typically results in the copyright being held by these institutions.
Furthermore, the decisions of whether to allow open-source distribution of the
code and the choice of licensing terms might be the prerogative of a university
representative (e.g., a PhD supervisor).
While this is in no way unique to non-CS domains, it is likely that pre-existing IP procedures are less likely to cover software dissemination aspects in institutions not dealing with computer science.
Even if the legal status of the code developed for a given experiment is settled and matches reproducibility requirements, the environment needed to run it might prevent unconstrained recomputation.
This in fact was also the case in one of the programs in question, as the code relied on a proprietary software library.

We also noted a lack of domain-specific workflows for recomputation in non-CS journals.
A counterexample is the Geoscientific Model Development (GMD) journal, which encourages reviewers to get
  acquainted with the code behind papers under review~\cite{GMD_editorial_2013}.
Yet, the same journal, as of now, imposes a 50~MB limit on the size of electronic supplements 
  which effectively rules out shipping a VM together with the paper.

The level of computer proficiency in non-CS domains~\cite{Merali_2010}
is also likely to influence the ability of researchers to use the VMs -- it is arguably less likely that a physics or urban-planning journal reviewer will be comfortable using a VM, in contrast to CS-related journals.

\section{Case study \#4: Recomputability of HCI Studies}
\label{s:group4}

Research in computer science often involves humans, especially if the subject of study is how people and machines interact with each other (e.g., in HCI). This kind of research focuses on phenomena that involve human agents and are therefore not addressable strictly through computation. However, there are also strictly computational issues that are critical for the replicability of HCI research. In this case study we looked at current practice in HCI research through an example published experiment, and aimed at identifying the road blocks and difficulties that present themselves when reproducing statistical analysis of data obtained from human behavior.  

\groupsubsec{7.1 Background}
The field of HCI heavily relies on the execution and analysis of empirical studies that involve humans. These results are used to, for example, build new interaction techniques and devices~\cite{Nacenta:2008}, and propose new models of human behavior with computers~\cite{Shoemaker:2012}. The reliability of the analysis and conclusions of these studies have been recently questioned. Problems identified include the fact that barely any results are replicated~\cite{hornbaek:replications}, that the research is often difficult or impossible to replicate~\cite{wilson:2011}, and that replicated results are difficult to publish and share to the larger community~\cite{wilson:2012}. The HCI community is currently trying to address some of these problems through new venues for publication (e.g., RepliCHI~\cite{wilson:2013,wilson:2014}), the creation of new tools~\cite{Mackay:2007}, and efforts to change the research culture and incentives (ACM CHI, arguably the most important conference in the field, introduced in 2013 a replication award or distinction for papers that address replicability).

Although the problem of replicability of experiments with humans is difficult and will likely require significant efforts from the community, the recomputability of these results and the associated statistical analysis has received relatively little attention, even though it is probably one of the most significant sources of inaccuracy and incorrect data in the field. Recomputability in HCI experiments refers mostly to the ability of others (not authors) to replicate the statistical analysis and statistical conclusions of a paper utilizing the same recorded data from the experiment. 

Quantitative experiments in HCI analyse data that is obtained from humans to draw conclusions that are relevant for the understanding or development of interfaces. Although the experiments are necessarily affected by the inherent variability introduced by humans, the analyses should not. Ideally, every researcher in the area, and more specifically, every reviewer of a paper containing statistical analysis of quantitative human data should be able to reproduce the analysis. The ability of reviewers to determine if a statistical analysis and interpretation of the data in a paper are correct is currently limited to checking that the reported degrees of freedom in an ANOVA (or a similar inferential statistic procedure) are consistent with the design of the experiment, and that the intermediate statistic figures (e.g., F values, DOF, p-values) are consistent with the statistic analysis. This is obviously not sufficient to detect even relatively simple errors during analysis that could mean the difference between radically opposite interpretations of the data. Examples that have been encountered by some of the authors of this paper include: reading statistics and degrees of freedom from an incorrect column in the software, reading statistics and degrees of freedom from an incorrect table, and performing within-subject analysis on between-subjects data. Some of these errors are virtually impossible to detect if the only provided information are the statistical figures typically found in papers. The problem is further magnified if the analysis is not standard. For example, if a new computational measure is created from the data, it might be impossible to reproduce without having the exact code, and if the analysis applies a machine learning approach there might be large numbers of parameters to adjust and many differently implemented variants of the same analysis (different analysis frameworks might have implementations of the same analysis that might lead to different results).

In order to prevent those errors and the significant loss of credibility of the data that they cause, authors should enable the recomputation of statistical and machine learning analysis on the data of any experiment, to the extent allowed by other ethics and privacy issues (Section~\ref{s:group1}). This requires that: a) authors make the data available, b) authors provide suitable meta-data that describes the semantics and structure of the data, c) authors provide instruction on how to reproduce the analysis. Sharing the data and the procedures of the analysis has advantages that go beyond the pure verifiability of the correctness of the result: the data can also be reanalysed (individually or in combination with other sources) to discover new insights, the analysis can serve as educational material for students in the area, and scientific fraud becomes, at least in theory, much harder to perpetrate. 

In this spirit of openness and scientific integrity, one of the authors (M.A.N.) has been striving to provide the data and the analyses for his own empirical research in HCI. Specifically, a recent project on the memorability of gestures~\cite{Nacenta:memorability} was developed from scratch as a pilot experience that would enable anyone to reproduce the analysis. For this purpose, the data and the basic analysis scripts necessary to perform the inferential statistics contained in the paper were prepared and included as an attachment to the original paper, which is currently accessible through the institutional research repository at the University of St Andrews~\cite{Nacenta:memorability_data}. This data and the required auxiliary files took approximately 6 hours to compile and prepare by the main author (excluding the time spent compiling and designing the statistical analyses). If this paper is representative of other work in the area, this amount of effort on the side of the authors does seem reasonable in exchange for the expected quality improvements for the field that recomputability could deliver. However, we have little knowledge about the challenges and difficulties encountered by the replicators (rather than the authors) in order to verify and check that the analysis is correct.

For this purpose, and in the context of the summer school that this article reflects on, we decided to set up an experiment in which the participants of the summer school (and authors of this article) with the exception of the author of the data, would try to replicate the results of the paper. The main objective of this research is to learn about the challenges and difficulties of a simple recomputation exercise of standard statistical analysis, to provide real examples of experience in recomputation of analysis in HCI, and to enable improvement of the provided data in the future.

\groupsubsec{7.2 Experience Report: Recomputing a Memorability Experiment}
The authors of this article (henceforth the \emph{reanalysts}), with the exception of
M.A.N. divided themselves into four teams (4, 3, 4, and 5 people per team), each of which would try to reproduce the same selection of results of the gesture memorability study reported in reference~\cite{Nacenta:memorability}. The target results for reanalysis were the averages and ANOVA analyses of the first paragraph of the /emph{Results} section of \emph{Experiment 3}. This paragraph contained three types of analysis: simple calculations of averages (recall rates), omnibus parametric ANOVA analyses, and pairwise post-hoc parametric t-tests. Approximately half of the reanalysts had a good understanding of HCI or had performed research in the HCI field. To provide sufficient background, the author of the reanalysed paper gave a 20-minute presentation on the content of the paper, aimed at a moderately knowledgeable audience. Reanalysts were allowed to ask any number of questions at the end.

The reanalysts received also a physical and a digital version of the original paper and a URL from where to download the data (as distributed originally to the public in~\cite{Nacenta:memorability_data}). The data is provided in a comma separated file (with column heading names in the first row). The data package also includes IBM SPSS Syntax files (SPSS's scripting language), and a README.txt file containing descriptions of the different files, including explanations of the columns. SPSS Syntax files were provided because it was the platform in which the analyses were performed, and it is commonly used as statistical software for the analysis of experiments in the HCI and Psychology communities. 

Teams were given approximately 90 minutes to replicate the results contained in the paragraph of the paper indicated above. Two groups opted to try to replicate the results by using SPSS (installed in the machines available to the reanalysts), one opted to replicate the results using R, and one opted to replicate the results using R while simultaneously recording the recomputation in a VM. The leader of the session provided help to the SPSS groups strictly on issues related to the SPSS interface. Each group was asked to assign one person to take notes on a paper notepad of the development of the session (specifically, steps taken, difficulties found, misunderstandings, and breakthroughs).

\groupsubsec{7.3 Results}
All the reanalysts spent the allocated time working on the recomputation of the results while recording on their notepads the actions and obstacles encountered. After the session was over, the reanalysts shared in public their results, conclusions and main obstacles for the benefit of the rest of the groups. The notepads were later analysed by M.A.N. by identifying problems, creating a physical affinity diagram of problems~\cite{hartson:2012}, and identifying the most relevant groups of related problems. The two following subsections report the degree of success achieved in the recomputation and the main categories of challenges and obstacles found.

\groupsubsubsec{7.3.1 Measures of Success}
Groups 2, 3 and 4 were able to achieve some verification of data present in the paper in the allotted time.

\emph{Group 1 (SPSS)} were able to open the data, read the README file, and run the script that loads and performs the analysis. They were not, however, able to find the appropriate correspondence between results on the paper and the output of SPSS. 
\emph{Group 2 (SPSS)} were able to open the data, read the README file, load the data with SPSS independently of the SPSS script, verify the integrity and structure of the data, run the scripts, find one of the ANOVA analyses in the output, and verify its correctness.
\emph{Group 3 (R)} were able to find and open the data, read the data with R, and run some basic descriptive statistics (averages).
\emph{Group 4 (R + VM)} were able to create a VM to store the analysis of the data, unpack the data, read the README file, failed at converting the provided SPSS Syntax scripts into R scripts, but were finally able to reproduce some basic descriptive statistics (averages).

\groupsubsubsec{7.3.2 Identified Problems and Challenges}
We identified four main groups of problems and challenges: tool problems, cross-tool problems, data and script problems, and lacks of knowledge.

\emph{Tool problems.} Reanalysts found it difficult to load data in SPSS, to run scripts, and and found the syntax scripts themselves non human-readable. The SPSS model of running scripts and presenting the results in very long report in a separate window/file was also found confusing. The SPSS Syntax Scripting facility is also difficult to get to work, and can be misleading (the system is not designed with the main goal of running full scripts), even for previous users of the tool. Additionally, the scripts cannot use relative file references, which forces the reanalysts to change the script itself instead of just running it (the folder structure of the reanalysis machine is not necessarily the same as the original machine where the data was first analysed). 

\emph{Cross-tool problems.} Reanalysts were unsure of whether the difference in
versions from the software used for the original analysis (SPSS 19), and the
tool available for reanalysis (SPSS 21) would cause problems. One group that
felt comfortable with R but wanted to take advantage of the provided SPSS Syntax
scripts tried to convert one to the other using an existing free
R package~\cite{spsstor}, however, the tool was found to be inadequate for this purpose; conversion from one language in one tool to another is a very complex problem, not likely to be solved soon. Additionally, the necessary installation of packages, dependencies, and the VM caused significant overhead.

\emph{Data and Script problems.} The data and scripts provided were also not ideal, and generated a number of problems and difficulties. Reanalysts detected inconsistencies in the naming of conditions and columns between the data and the paper, which are due to the authors of the original paper renaming conditions and columns to make the paper more readable. Some groups also tried to identify data based on the SPSS-generated graphics, but these do not correspond to the graphics used in the final version of the paper (SPSS graphics are not of the quality and format required in most scientific publications, and therefore had to be redone). This caused confusion to three groups. Finally, the analyses provided in the SPSS Syntax are exhaustive, containing much more information than the paper. This caused confusion in reanalysts, who had serious difficulties relating the output generated by the scripts with the data reported in the paper. This was sometimes made worse by the fact that the other was different in both systems.

\emph{Required Knowledge Breadth.} All groups highlighted the depth and breadth of knowledge required to achieve recomputation of data. At the lowest levels of abstraction, reanalysts had to be knowledgeable in SPSS operation. Knowledge on the use of data formats is also a requirement. Those groups that used R for analysis did not only have to show a significant mastery of R, but also of the relationship between R and SPSS Syntax and, more importantly, of the specific statistical procedures and how they are performed in both platforms. Finally, the reanalysts had to achieve a grounded understanding of the experimental design and purpose of the experiment, something that requires detailed and thoughtful study.

\groupsubsec{7.4 Discussion and Recommendations}
Although the main focus on recomputation in HCI has focused on the replication of empirical data collection (replicated experiments), there is still much to do (and much benefit to get) from improving the recomputability of the analyses of the data gathered. In this section we discuss the main issues and lessons learned from our experience, as well as limitations from our methods and a set of recommendations on how to improve the impact and feasibility of recomputation in HCI.

\groupsubsubsec{7.4.1 Reasonable Success}
Our experience shows that a group of motivated individuals achieved a modest amount of success in reanalyzing a set of simple statistical analysis of HCI empirical data. The results suggest that recomputability is within reach of the experimental HCI community, and data and analysis sharing practices will allow researchers with a stake in the correctness of other researchers' results to verify their analysis. This is possible even in the current state of affairs (many different tools being used, lack of explicit support for recomputation), but requires a significant amount of time, effort, and expertise from multiple sources. This effort and time is often not available for recomputation scenarios that require agile and fast reanalysis, such as paper article reviewing. For this, tool support and a culture change will be required. 

\groupsubsubsec{7.4.2 Tools are Key (and not ready)}
SPSS might be an adequate tool for performing statistical analysis; it is successfully used by many in HCI and many other areas. Reanalysis, however, imposes a different set of constraints and requirements, and our reanalysts had many problems with the tool. Some problems relate to the general usability of the tool (which makes reanalysis difficult if you are not an SPSS expert), some to the implicit design assumption in SPSS that the data is collected, analysed and interpreted by the same person. One of the key problems of using SPSS to enable recomputation is that there is no easy way to establish a clear correspondence between the results of the analysis in SPSS and the specific statistics extracted for the paper text, tables and graphics. 

R seems better suited for these tasks. It is possible to write R code that integrates with the text through Sweave~\cite{lmucs-papers:Leisch:2002} so that the specific analyses are compiled together with the PDF document. This makes the origin and procedure used to obtain a particular numerical result traceable to the data, and therefore easier to check and recompute. Although this is highly desirable it might still be unreasonable to demand that everyone writing or reviewing HCI and psychology papers master a programming language and tools that are generally not renowned for their usability, and that everyone is able to deal with the installation hassles of R, Sweave, ggplot2, \LaTeX, etc. in their operating system of choice. There is room for improvement for these tools, and distributing VMs may further help, but commercial tools still have an opportunity to retain their business if they provide features that adapt to the demands of easy recomputation and better support for scientific reporting. Although our experience only involved R and SPSS, the example extrapolates to other commercial tools (e.g., SAS) and open source projects (SciPy) in the statistical arena.

\groupsubsubsec{7.4.3 A Culture Change}
Recomputation is therefore feasible and likely to become easier in the near future through better support and tools. However, it is unclear whether the HCI research community will embrace it. Recomputability requires more work for researchers writing papers, new habits in the analysis and reporting of experiments and, for most researchers, learning and mastering new tools. A change of culture will, however, not only mean better science through more recomputable results, but also enhanced opportunities for new analysis on old data, enabling learning from others, and making scientific fraud and bad practices easier to detect. For this all to happen, we need to start demanding from authors that they share data and analyses, and that they consider the needs of the reanalyst while planning, performing, and reporting their quantitative empirical research. A small example of this are current efforts by one of the authors to make data available to the research community through purpose-made interfaces that enable analysis and reanalysis of previous results \cite{Grijincu:2014}. 

\groupsubsubsec{7.4.4 Limitations}
To our knowledge, this section reports the first study of recomputation of the statistical analysis of HCI empirical data. We have been able to learn valuable lessons from this experience, including ways to improve the actual data and analysis kit for the original study. However this only represents a semi-informal study with semi-controlled observation for one specific case analysed using a specific tool (SPSS). Further research is required to validate these results and generalize the lessons learned to other tools and other types of reanalysts; specifically, it would be useful to investigate how experts in a particular field go about reanalysing existing results, and what are the specific barriers present when the data and analyses are prepared with a more sophisticated system such as R with Sweave and \LaTeX.

\groupsubsubsec{7.4.5 Recommendations}
For the recomputability of quantitative analyses in HCI research, we make the following recommendations:
\begin{itemize}
\item When possible, share the raw data and analysis for experiments to enable recomputability.
\item Aim to reduce the knowledge required to reanalyse data. Reanalyst teams already require knowledge of the topic area, the reanalysis tool, and the computational procedures.
\item Make results explicitly traceable from computation to report.
\item After the paper is written, revise and adapt data for consistency of nomenclature of factors and condition names.
\item Due to cost, fitness and availability, favor open source tool platforms for analysis and reanalysis preparation (at least for the moment).
\item To reduce overheads due to learning of open tools by reanalysts, provide also clear instructions with the data and links to resources for learning and using the reanalysis tools.
\item To establish a replicability research culture, demand that research authors provide data and analysis at publication time.
\end{itemize}

\section{Open, Executable, Reproducible}

This paper is intended to be \emph{open}, 
\emph{executable},
and \emph{reproducible}.  
We discuss here what we mean by these words, how we tried to achieve them, and to what extent we succeeded.

By ``open'', we mean that the paper was openly developed and written.  The paper was first sketched out in the week of the summer school, and then 
developed further over the following three weeks.  The authors collaborated via GitHub: this is not unusual but distributed source code control was  particularly important with so may authors working in parallel over a short period of time.
Because we used a public GitHub repository~\cite{summerschoolpaper}, 
the
development of the paper can be tracked throughout through the commit history on
GitHub.  
Anybody can download not only the final paper, but its source, and code and other materials collected during its development.  
An interesting aside, relevant to cases where statements of contribution are required, is that anybody can see what each author committed on the paper or supporting materials throughout its life.  It must be borne in mind, however, that a commit by one person may represent the work of several authors working together offline.  

By ``executable'', we mean that the paper that can be reconstructed from source 
materials, and that data can be reanalysed as it changes and new versions of the paper produced, and possibly executable code rerun.   The name has been used, for example, by the ``Executable Papers Grand Challenge''~\cite{executablepaper}.
This has numerous advantages because as we add data, the paper does not need to be rewritten.\footnote{The name ``reproducible paper'' is sometimes used for this, but can lead to confusion because a paper can be executable in the sense of being able to produce new figures with 
changed data, but not reproducible if that data can not be reconstructed ab initio.}
To make our paper executable, we used Sweave~\cite{lmucs-papers:Leisch:2002}, a package that integrates the statistics system R and \LaTeX. 
To ease the workflow and to make execution of the paper easy, we wrote a Makefile for the generation of the paper PDF, although there are still some issues which need manual intervention such as installation of the necessary R packages. 

By ``reproducible'', we mean that it is our intention that other scientists (or
ourselves at later dates) will be able to reproduce our work to assess if
statements we make are correct and if conclusions are valid.  
As mentioned earlier, 
To enable this we
have attempted to collate materials necessary for each study, and make them
available to future researchers.  
In some cases this has also been done in git,
with materials such as ethics forms and experimental results put into the repository.  
In other cases we have constructed VMs to recompute experiments.  
We also built a machine which not only contains a clone of the GitHub
repository, but all the necessary software and packages, such as R and
\LaTeX, to build the paper. We make this available on both
recomputation.org\footnote{\url{http://recomputation.org/emcsr2014/}}
and the Microsoft VM
Depot.\footnote{\url{http://vmdepot.msopentech.com/Vhd/Show?vhdId=44582}}
As well as reproducibility, this helped us during the writing of the paper.  At times some authors would be unable to build the whole paper, perhaps because of slight differences in version numbers of R or its packages from other authors.  The availability of the VM in which the paper could build was invaluable, since if it built there, we were safe.  This also highlights the value of the paper VM, since if not all authors could build the paper during its preparation because of package inconsistencies, it is very likely that future workers would not be able to build it from the GitHub repository without some work.
Even a small VM 
might be half a gigabyte, and such large files can be problematic for git.
These are therefore stored elsewhere.  
Our approach can be seen as similar to that of Brown~\cite{brown} for one of his papers with colleagues~\cite{Brown2012}.

To what extent have we succeeded in our efforts?
Our success is mixed.  We cannot be completely open because
some of the data and/or programs used in various parts of our paper do not allow us to share them.  This is particularly true of the third case study, and as a result we cannot distribute the VM embodying those experiments. Also our paper is not fully executable since 
many computations involved in constructing the data must be run by hand.  Also some of the tables in this paper are static, with data entered manually rather than reanalysed by Sweave. This contrasts negatively with Brown's executable paper~\cite{brown}.  
In terms of reproducibility, we feel we have been mostly successful.  
Provision of materials via GitHub will be helpful for those who wish to adapt or reproduce our work.
We have provided VMs where possible and appropriate, and can distribute two of them: this should enable the recomputation of our paper.  This can be performed either locally, or using VMs deployed in Microsoft Azure requiring no software installation on a user's machine. 
While we might pat ourselves on the back for this, we feel it is better to be cautious. We ourselves have discussed potential issues, such as ethical and legal, and unforeseen technical issues might prevent reproducibility. The real test of reproducibility must be time.
It is interesting to speculate: for example, if we run a second summer school in 2015, how hard would it be for participants 
to reproduce this paper?

\section{Discussion and Conclusions}
\label{s:discussion}

The experiences we have reported on in this paper cannot be taken as full scale studies from which conclusions can definitely be drawn, but aim to provide evidence and some systematic reflection of the current state of scientific reproducibility as seen from the researchers' viewpoint.  For example, the 
selection of ethics forms we obtained was certainly subject to selection bias: they were typically the ethics forms that were easy to obtain.  So we cannot state hypotheses and say our analyses definitively show the hypothesis is supported.   Having given that caveat, we feel that our investigations have uncovered a number of interesting points.  The remarkable diversity of ethics forms enforced around the world raises serious issues for reproducibility.  It also suggests an unnecessary cause of inertia in reducing the speed with which reproductions of experiments can be started up.  Our parallel experiments suggest that the difficulties of reproducibility are 
multiplied by the extra complications inherent in this domain.  Our
recomputation of non-computer-science experiments was encouraging in
that it was achieved in a relatively short amount of time.
Nevertheless important issues arose, such as the impossibility of
distributing the resulting VM because of legal issues. 
Such issues need attention and easing as much as possible.  In many
cases literally nobody knows what the answers to some of the legal
questions are: this is not because they are necessary complex but
because they have not been specifically assessed by lawyers, or
ultimately tested in court.  This means that we have a situation where
we suspect that how we are reproducing work is legal, but do not know
it is: a position that might be comfortable in quantum mechanics but
less so to the more classically binary computer scientist. Finally,
our attempt at recomputing an HCI experiment that was thought to be
reproducible was partially successful, but more importantly has raised
a useful set of recommendations and lessons for future recomputable
research.

\section*{Acknowledgements}
\label{s:ack}

The authors of this paper include organisers and participants in the
summer school. We are very grateful to all of the people who helped
make the school a success and want to acknowledge everybody who is not an
author.

First we thank the organisers: John McDermott, Angela Miguel, and Lakshitha de Silva for organisational and technical help.

We thank the various sponsors of the summer school for financial and
other forms of assistance:
SICSA (the Scottish Informatics and Computer Science Alliance); 
Microsoft Research; the Software Sustainability Institute; 
the St Andrews School of Computer Science;
and the EPSRC Impact Acceleration Award ``Recomputation.org: Making
Computational Experiments Recomputable''
at the University of St Andrews.

We are also grateful to speakers at the summer school who were not also authors of this paper: 
Neil Chue Hong,
Stephen Crouch, 
Darren Kidney, and
Burkhard Schafer.

\bibliographystyle{abbrvDOI}
\bibliography{paper}

\end{document}